\documentclass[12pt,titlepage]{article}%
\usepackage{graphicx}
\usepackage{amsmath}
\usepackage{amsfonts}
\usepackage{amssymb}
\usepackage{color}%
\usepackage[font=small,labelfont=bf,width=\textwidth]{caption}
\providecommand{\U}[1]{\protect\rule{.1in}{.1in}}
\begin{document}

\title{Optimal Initiation of a GLWB in a Variable Annuity: No Arbitrage Approach}
\author{H. Huang\thanks{Huang is Professor of Mathematics and
Statistics at York University. Milevsky is Associate Professor of Finance,
York University, and Executive Director of the IFID Centre. Salisbury is
Professor of Mathematics and Statistics at York University, all in Toronto,
Canada. The contact author (Milevsky) can be reached via email at:
milevsky@yorku.ca. The authors acknowledge the helpful comments from participants to the IFID Centre conference. Huang's and
Salisbury's research is supported in part by NSERC and MPRIME. }, M. A. Milevsky and T.S. Salisbury}
\date{Version: 25 February 2013}

\maketitle

\begin{abstract} 

\begin{center}
\emph{OPTIMAL INITIATION OF a GLWB in a VARIABLE ANNUITY: \\ NO ARBITRAGE APPROACH}\medskip
\end{center}

This paper offers a financial economic perspective on the optimal time (and age) at which the owner of a Variable Annuity (VA) policy with a Guaranteed Living Withdrawal Benefit (GLWB) rider should initiate guaranteed lifetime income payments. We abstract from utility, bequest and consumption preference issues by treating the VA as liquid and tradable. This allows us to use an American option pricing framework to derive a so-called optimal initiation region. Our main practical finding is that given current design parameters in which volatility (asset allocation) is restricted to less than $20\%$, while guaranteed payout rates (GPR) as well as bonus (roll-up) rates are less than 5\%, GLWBs that are in-the-money should be turned on by the late 50s and certainly the early 60s. The exception to the rule is when a non-constant GPR is about to increase (soon) to a higher age band, in which case the optimal policy is to wait until the new GPR is hit and then initiate immediately.  Also, to offer a different perspective, we invert the model and solve for the bonus (roll-up) rate that is required to justify delaying initiation at any age. We find that the required bonus is quite high and more than what is currently promised by existing products. Our methodology and results should be of interest to researchers as well as to the individuals that collectively have over \$1 USD trillion in aggregate invested in these products. We conclude by suggesting that much of the non-initiation at older age is irrational (which obviously benefits the insurance industry.)

\end{abstract}

\section{Introduction}
\label{intro}

From the point of view of an insurance company issuing or selling a \emph{Variable Annuity with a Guaranteed Lifetime Withdrawal Benefit} (from here: GLWB), natural questions are the pricing, hedging, and risk management of this product. From the point of view of the purchaser, a different set of questions are important -- optimal product allocation to GLWB's, optimal asset allocation within GLWB's, optimal management of multiple GLWB accounts (eg. where and when should new deposits be made), and optimal initiation of withdrawals. These are all mostly open and ongoing research issues, and the latter is the question considered here.

A GLWB purchaser decides when to initiate withdrawals. Typically, the vendor rewards the purchaser for delaying initiation by providing a bonus (a.k.a. roll-up) to the guarantee base of the product. The guarantee base may also rise because of resets, which are more likely if initiation is delayed. Once initiated, payments are made at a rate which is a percentage of the guarantee base. This percentage may vary depending on the age at initiation. Insurance fees for the GLWB are charged on the guarantee base, regardless of whether withdrawals have been initiated. At the time of initiation the bonus (roll-up) ceases but the guaranteed withdrawals do not trigger surrender charges. We do not model withdrawals that exceed the guaranteed amount, since in that case we would anticipate lapsation of the entire GLWB. That would be a reasonable alternative if we were studying whether or not the GLWB is itself a suitable investment. Instead we have in mind a situation in which a firm decision has been made to retain the GLWB (perhaps due to adverse tax implications) and the question of interest is purely one of initiating versus delaying withdrawals. We anticipate that this is a scenario faced by many individuls who purchased variable annuities with GLWBs in the last few years; currently `under-water' (ie. the account value is under the guaranteed withdrawal base.) In the industry's language, should the investor stop the accumulation process and begin withdrawals? Or, should they continue accumulating? We note that the typical buyer of a GLWB (in the U.S.) is in his or her mid 50s, or early 60s, and that there are no additional tax penalties imposed on withdrawals after the age of 60.

One way of posing the question of the optimal initiation time is to ask what scenario the issuer must plan for in order to be fully hedged. In other words, what is the most costly scenario from the point of view of the issuer? Naturally this is also the optimal initiation time, from the point of view of a hedge fund that has purchased multiple VA contracts from the original clients (enough to be diversified from the point of view of mortality risk). We call this the {\it risk-neutral initiation problem}. A nice feature of this formulation is that there is a unique and easily understood answer to the question, based purely on a complete-market economic analysis. Auxiliary assets (eg. holdings in a non-guaranteed account) are irrelevant. We present this analysis in this paper. 

A different point of view is what we might call the {\it utility maximizing initiation problem}, namely that of a purchaser who genuinely wishes to hold the longevity protection offered by a VA. There are a variety of possible motivations for purchasing the VA, and optimal behavior may in principal differ depending on the actual goal of the purchase. Put another way, purchasers may have a diversity of utility functions, that typically blend some combination of lifetime consumption and preservation of capital or bequest. This approach might lead to results that are quite different from a No Arbitrage analaysis. Nevertheless, preliminary results (not presented here) are entirely consistent with the conclusion from the risk-neutral version of the problem: Given current product features, typically it is optimal to initiate immediately, exceptions being for particularly young individuals, individuals within a short time of a rise in withdrawal rates, or individuals holding products with extreme return  characteristics (eg very high volatility or bonus rates).

The remainder of this paper is organized as follows. In Section \#2 we briefly describe the existing literature on variable annuity (VA) guarantees to help position our contribution within the literature. Then, in Section \#3 we describe the risk-neutral model in which we operate. Section \#4 provides the numerical examples and illustrations and Section \#5 concludes the paper. All figures and tables that are referenced appear after the bibliography.

\section{Position within the Literature}

Scholarly research into the area of variable annuities (VAs), and specifically guaranteed minimum benefits (GMXBs) options -- has experienced much growth in the last decade or so, and a sizeable portion of this work has been published in the IME. One of earliest papers analyzing options inside VAs and their Canadian counter-part called Segregated Mutual Funds, was Windcliff, Forsyth and Vetzal (2001). They analyzed the ‘shout option’ while Milevsky and Posner (2001) examined the ‘titanic option.’ To our knowledge, the first formal analysis of the guaranteed minimum withdrawal benefit (GMWB) was Milevsky and Salisbury (2006), which focused on the cost of hedging a promise to provide a fixed-term annuity. The main conclusion of Milevsky and Salisbury (2006) was that GLWBs appeared to be underpriced, relative to what was being charged for them in the (US) market. This result was echoed by Coleman, Li and Patron (2006) as well as Chen, Vetzal and Forsyth (2008), who also obtained option values. Although they employed a different PDE-based methodology to derive hedging costs under a variety of parameter values, they arrived at similarly high costs. The work of Dai, Kwok and Zong (2008) further re-enforced that fact that U.S. insurance companies were not charging enough for guaranteed living benefit riders on Variable Annuities. In fact, some might argue that this result was foreshadowed by Boyle and Hardy (2003), who examined the valuation of (UK-based) guaranteed annuity options (GOA), which are somewhat different from GLWBs or GMWBs, but arrived at similar conclusions. The insurance industry was under-pricing, under-reserving and mis-hedging these complex options.

During and after the global financial crises of 2007 and 2008, most insurance companies in the U.S. and Canada finally realized this and pulled-back on their (aggressive) offerings. It is very difficult to find VAs with benefits and features such as the ones described in the above-cited research papers. But, those investors fortunate enough to have purchased these product prior to their withdrawal from the market must now decide how to optimize their own withdrawals from the product.

As far as more recent – post crises research – is concerned, the earlier pricing and valuation results have been extended to include more complex models for asset returns, mortality risk as well as policyholder behavior. Of course, none of these papers have invalidated the result that these options are quite  complicated and rather valuable to the consumer. If anything, they prove that the options are even more complicated than previously thought. For example, Ng and Li (2011) use a multivariate valuation framework and so-called regime switching Esscher transforms to value VA guarantees. Feng and Volkmer (2012) propose alternative analytical methods to calculate risk measures, which obviously reduce computational time and is prized by practitioners. In a sweeping paper, Bacinello, Millossovich Olivieri and Pitacco (2012) offer a unifying valuation framework for all guaranteed – living and death – benefits within variable annuities. The above papers and other extensions by the same authors in different venues all focus on the financial market uncertainty vis a vis the optionality, more so than mortality and longevity uncertainty. The paper by Ngai and Sherris (2011), for example, focuses on longevity risk and the effectiveness of hedging using longevity bonds, but doesn't discuss the optionality and timing problem within variable annuities.

In conclusion, most of the above articles are concerned primarily with risk management from the issuers point of view, and aren't concerned with the normative or prescriptive implications for individuals who seek to maximize the embedded option value. An exception to this is the recent paper by Gao and Ulm (2012) who offer suggestions on the optimal asset allocation within a VA with a guaranteed death benefit. Their conclusion is that the `Merton Ratio' asset allocation percentage doesn’t necessarily hold within a VA, because of the guaranteed death benefit. Our paper is in the same individual-focused vein, which we believe is a very important, but often a neglected aspect of the research on risk management and valuation.

In this paper we focus exclusively on the problem from the perspective of the individual (retiree) who seeks guidance on when to initiate or begin withdrawals from the guaranteed living withdrawal benefit (GLWB). In particular we are interested in the optimal timing of annuitization, similar in spirit to the work by Milevsky and Young (2008) or Stabile (2006), which solves an optimal timing problem by formulating and solving the relevant HJB equation. This isn't quite annuitization in the irreversible sense, given the liquidity of the account. But the idea is the same. 

As discussed in the introduction, by adopting a No Arbitrage perspective we are assuming the individual is trying to maximize the cost of the guarantee to the insurance company offering the GLWB. The optimal policy is the one that is the most costly to the issuer. Preliminary results (not reported) support the conclusion that this is optimal even under a utility-based analysis. Something which is properly left for further research. In either case, the practical consequences (for the individual) are that given the age (above late 50s) of typical investors holding substantial GLWB's, current (low 3\%) interest rates, existing (low 5\%) bonus rates, forced (low 15\%) volatility allocation, low guaranteed payout rates (4\% to 5\%), it is optimal to initiate the income immediately. The next section describes the set-up.

\section{The risk-neutral initiation problem}
\label{rnanalysis}

We start by giving the optimal initiation time, in the sense that it is the most costly for the insurer, or the most rewarding for a (diversified) hedge-fund investor who has bought the VA contracts. We will ignore lapsation, will assume that all contracts purchased have identical returns and initial deposits of $M_0=1$, that all are sold to clients of the same age and health, that they initiate  at the same time, and will take the number of contracts so large that mortality is completely diversified. And, while stochastic mortality (as well as stochastic interest rates and stochastic volatility) are all fertile areas of research, we start with a simple model to gain clear intuition about the parameters and factors that drive the initiation decision.

Our models are developed in a continuous-time (stochastic calculus) framework in which step-ups, roll-ups, interest rates and investment returns accrue continuously. This is obviously done for analytic convenience (and academic tradition). However, in a later section we conduct and report on Monte Carlo (MC) simulations that assume annual accruals and compare results to the PDE-based approach, in Table \#1. We find that numerical results and specifically the initiation regions, as well as the required bonus rates to justify the delay of initiation, are quite similar.

\subsection{Notation and dynamics}
\label{sec:rnnotation}

We define the following quantities:
\begin{alignat*}{2}
X_t&=\text{VA account value} &
M_t&=\text{guarantee base ($M_0=1$)}\\
Y_t&=\frac{X_t}{M_t}=\text{moneyness} &
L_t&=\text{Local time of $Y_t$ at the level $y=1$}\\
N_t&=\text{Proportion of initial clients still alive}\quad&
V_t&=\text{Total value of the hedge}\\
C&=\text{Total initial assets deposited}&
\lambda_t&=\text{mortality hazard rate}\\
\tau&=\text{initiation time}&
R&=\text{Ruin time (so $X_R=0$)}\\
r&=\text{risk-free growth rate}&
\beta&=\text{bonus rate}\\
g_\tau&=\text{payout rate for initiation at time $\tau$}\quad &
\alpha&=\text{insurance fee as percent of $M_t$}\\
\sigma&=\text{VA account volatility}&&
\end{alignat*}
We use risk-neutral GBM dynamics, with continuous stepups, bonuses, and fees. So prior to ruin we have
\begin{align*}
dX_t&=r X_t\,dt+\sigma X_t\,dB_t-M_t(g_\tau1_{\{\tau<t\}}+\alpha)\,dt\\
dM_t&=M_t\,dL_t+\beta M_t1_{\{t<\tau\}}\,dt
\end{align*}
which implies that prior to ruin ($t<R$),
$$
dY_t=\frac{dX_t}{M_t}-\frac{X_t\,dM_t}{M_t^2}=(r-\beta 1_{\{t<\tau\}}) Y_t\,dt+\sigma Y_t\,dB_t-(g_\tau1_{\{\tau<t\}}+\alpha)\,dt-dL_t.
$$
This is a Skorokhod-type equation, in which the $dL_t$ term keeps $Y_t\le 1$, ie $X_t\le M_t$.
Ruin is unlikely prior to initiation, but is theoretically possible because of the fee structure. Once the account ruins, clients are obliged to initiate withdrawals (and there certainly is no incentive to delay any further.) Now, because mortality is diversified, 
$$
dN_t=-\lambda_t N_t\,dt.
$$
We ignore non-hedging fees, so fees collected stay in the hedge, and cash flows derive from withdrawals and refunding account balances (albeit with no GMDB) at death. Other than those cash-flows, the hedging portfolio is self-financing, therefore
\begin{equation}
dV_t=rV_t\,dt-CN_tM_tg_\tau1_{\{\tau<t\}}\,dt+CM_tY_t\,dN_t+d\text{Mar}_t,
\label{eqn:cashflow}
\end{equation}
where $\text{Mar}_t$ is a martingale.

\subsection{Hedging}
\label{sec:hedging}
Our goal is to choose $\tau$ to maximize the $V_0$ required to ensure that $V_t\ge 0$ for all $t$. 
By scaling, and the fact that the $Y_t$ dynamics do not depend on $M_t$, we may write 
\begin{equation}
\label{eqn:valuefunction}
V_t=CM_tN_tv(t,\tau,Y_t),
\end{equation} 
where $v$ denotes the hedge value required per dollar of remaining guarantee. 
There are, of course, several versions of this, depending on whether ruin or initiation has occurred.  In the following, we make our variables correspond to $Y_t=y$ and, if initiation has already occurred, also $\tau=s\le t$. 
\begin{align*}
v^0(t,y):&\quad\text{value function before initiation or ruin, }\\
v^1(t,s,y):&\quad\text{value function after initiation but before ruin,  }\\
v^2(t,s):&\quad\text{value function after initiation and ruin.  }
\end{align*}

\subsection{The post-ruin case $v^2$}
\label{sec:v2regimern}
At ruin time clients initiate whether they have done so before or not. Therefore we are dealing with an annuity, though the payout rate depends on the age of the client at the time of initiation. In other words, 
$$
v^2(t,s)=g_s\int_t^\infty e^{-\int_t^q(r+\lambda_p)\,dp}\,dq=g_s\bar{a}_t
$$ 
where $\bar{a}_t$ is the price of an annuity paying $\$1$ for life, starting at time $t$. $v^2$ will appear as a boundary condition below.

Alternatively, we can state this in differential form, to make clear the relation with $v^1$ and $v^0$. Since $M_t$ no longer changes, we may apply Ito's lemma to \eqref{eqn:valuefunction} and match terms in \eqref{eqn:cashflow}, or simply differentiate $v^2$ with respect to $t$. This leaves us with the ODE
\begin{equation}
\label{v2pde}
v_t^2-(r+\lambda_t) v^2=-g_s,
\end{equation}
with boundary condition $v^2(\infty,s)=0$ because we assume the hazard rate increases without bound. The martingale term vanishes.

\subsection{The post-initiation, pre-ruin case $v^1$}
\label{sec:v1regimern}
Apply Ito's lemma to \eqref{eqn:valuefunction} and match terms in \eqref{eqn:cashflow}, we get
\begin{multline*}
CM_tN_t\Big[v_t^1\,dt+v_y^1(rY_t\,dt+\sigma Y_t\,dB_t-(g_\tau+\alpha)\,dt-dL_t)+\frac12 \sigma^2 Y_t^2v_{yy}^1\,dt+v^1\,dL_t-\lambda_t v^1\,dt\Big]\\
=CM_tN_t\Big[rv^1\,dt-g_\tau\,dt-\lambda_tY_t\,dt\Big]+d\text{Mar}_t.
\end{multline*}
For $s$ fixed, this yields the following PDE in the two variables $t>s$ and $0<y<1$:
\begin{equation}
\label{v1pde}
v_t^1+\big[ry-(g_s+\alpha)\big]v_y^1+\frac12 \sigma^2 y^2v_{yy}^1-(r+\lambda_t )v^1
=-g_s-\lambda_ty.
\end{equation}
The boundary conditions are $v^1(t,s,1)=v_y^1(t,s,1)$ (from the $dL_t$ terms), that $v^1(t,s,0)=v^2(t,s)$, and $v^1(\infty,s,y)=y$ (since the remaining cohort dies immediately). 

\subsection{The pre-initiation case $v^0$}
\label{sec:v0regimern}
In the continuation region, the same analysis as above gives that 
\begin{multline*}
CM_tN_t\Big[v_t^0\,dt+v_y^0((r-\beta)Y_t\,dt+\sigma Y_t\,dB_t-\alpha\,dt-dL_t)+\frac12 \sigma^2 Y_t^2v_{yy}^0\,dt+v^0\,dL_t+(\beta-\lambda_t) v^0\,dt\Big]\\
=CM_tN_t\Big[rv^0\,dt-\lambda_tY_t\,dt\Big]+d\text{Mar}_t.
\end{multline*}
This yields the following PDE in the two variables $t>0$ and $0<y<1$:
\begin{equation}
\label{v0pde}
v_t^0+\big[(r-\beta)y-\alpha\big]v_y^0+\frac12 \sigma^2 y^2v_{yy}^0-(r+\lambda_t -\beta)v^0
=-\lambda_ty.
\end{equation}
The boundary conditions are $v^0(t,1)=v_y^0(t,1)$ (from the $dL_t$ terms), $v^0(\infty,y)=y$ (since the remaining cohort dies immediately), and on the free boundary we have $v^0(t,y)=v^1(t,t,y)$ and $v^0_y(t,y)=v^1_y(t,t,y)$ (smooth pasting). Because initiation is forced upon ruin, this in principle could include the condition $v^0(t,0)=v^2(t,t)=g_t\bar{a}_t$.

\subsection{Solution}

It would naturally be of interest to evaluate the actual initial hedging cost $v^0(0,1)$, and to determine if it is $>1$ or $<1$. In case $v^0(0,1)>1$, the issuer is not able to fully hedge. We do not focus on this, however, since the question of  interest to us is the initiation time (ie the shape of the initiation boundary), not the hedgeability of the product itself. In other words, we do not opine on whether $\alpha$ is large enough to cover the hedging cost.

We solve numerically, by discretizing $s$, $t$, and $y$. In full mathematical generality, $g_s$ could vary continuously with $s$, in which case we would proceed as follows. Fix successive grid points $t_k<t_{k+1}$. Assume that we have calculated $v^0(t,y)$ and $v^1(t,s,y)$ for all grid points $0\le y\le 1$ and $t\ge s\ge t_{k+1}$. To  obtain the new values $v^0(t_k,y)$ we solve the entire PDE \eqref{v1pde} in $(t,y)$ for $s=t_k$, $t>t_k$ and $0<y<1$. We then calculate a delay (continuation) value $\tilde v^0(t_k,y)$ from one step of the PDE \eqref{v0pde}, and set $v^0(t_k,y)=\max(\tilde v^0(t_k,y),v^1(t_k,t_k,y))$. Clearly the time-consuming portion of this calculation is the need to re-solve the PDE \eqref{v1pde} at every step.

In practice, this issue does not arise, as real payout schedules $g_s$ do not step up continuously. Typically they step up at a small number (3-4) of specified age bands $s_k$. This greatly simplifies the computation, as we may carry out a version of the above procedure with just these values $s_k$ for $s$, and then varying only $t_i$ and $y_j$ on a finer grid. 

Note that numerical computation of  the PDE solution $v^0$ is not the only possible approach to this problem. Recall that $N_0=M_0=1$. So by \eqref{eqn:cashflow}, we may write $v^0(0,y)$ as the risk neutral expectation of discounted cash flows from the hedge:
$$
v^0(0,y)=\sup_\tau E_Q\Big[\int_0^R e^{-rt}M_tY_t\,d(-N_t)+\int_\tau^\infty N_tM_tg_\tau\,dt\mid Y_0=y\Big].
$$
As alluded to earlier, Monte Carlo (MC) methods could in principal be used to calculate $v^0$, except that the optimization over $\tau$ is hard to implement using simulation. However, the approach does give us a simple way of comparing our results with specific initiation strategies $\tau$, to see how close they are to being optimal. Each such strategy naturally provides a lower bound for $v^0$. We can easily gauge other effects using MC, for example, the impact of using annual stepups, bonuses, or fees, instead of the continuous versions implemented in the PDE approach. We will make such comparisons in the following section, where we compare two strategies: initiate now, versus initiate in 5 years. We will compute the ``break-even'' bonus rate, ie the $\beta$ at which one is indifferent between the two strategies. 
We do so using the MC approach (stepup/bonus/fee adjustments made once per year), and compare the results obtained using the PDE approach (continuous stepups/bonuses/fees) that gives the corresponding bonus rate based on the optimal strategy with a five-year waiting period.

\section{Results and discussion}

\begin{figure}[here]
\vspace{-2in}
\begin{center}
\includegraphics[width=0.8\textwidth]{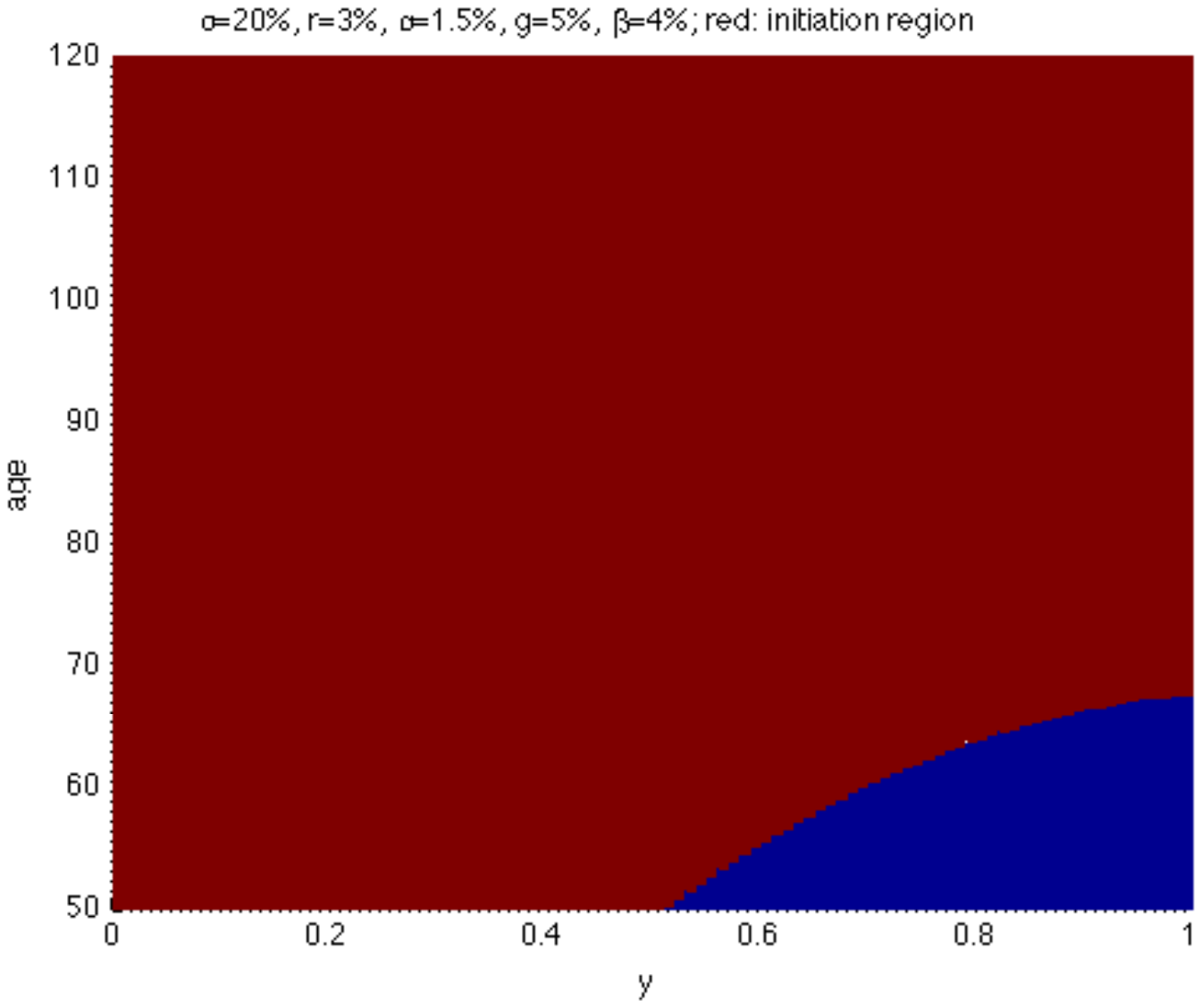} 
\vspace{-2in}
\caption{(Moderate $\sigma$) Model assumptions: $r=3\%$ (=interest rate), $\beta=4\%$ (=bonus rate), $m=87.25$ (=Gompertz modal value), $b=9.5$ (=Gompertz dispersion value), $50\le\text{ age }\le120$, $\sigma=20\%$ (=account volatility), $\alpha=1.5\%$ (=GLWB insurance fees), and $g=5\%$ (=guaranteed payout rate). The red region denotes the combination of money-ness and age in which accumulation phase should be terminated and the GLWB should be initiated. The green region is where the option to wait is valuable.}
\label{gcont}
\end{center}
\end{figure}

When the guaranteed payout rate (GPR) $g$ is constant and independent of age, there is typically a single contiguous initiation region, as shown in Figure \ref{gcont}, for example. There the parameter values are chosen to be $r=3\%$, $\beta=4\%$, Gompertz mortality parameters $m=87.25$, $b=9.5$, $50\le\text{ age }\le120$, $\sigma=20\%$, $\alpha=1.5\%$, and $g=5\%$. The high upper-age horizon of 120 is chosen as the maximum for life. The red-colour area denotes the region in which the accumulation should be terminated and the GLWB should be initiated. The green-coloured region denotes the age and money-ness in which accumulation should continue and withdrawals should not be initiated. Recall that it is only during the accumulation period that a bonus (roll-up) will be credited to the guaranteed base.

\begin{center}
\bf{Figure \#1 to \#5 Here}
\end{center}

\begin{figure}[here]
\vspace{-2in}
\begin{center}
\includegraphics[width=0.8\textwidth]{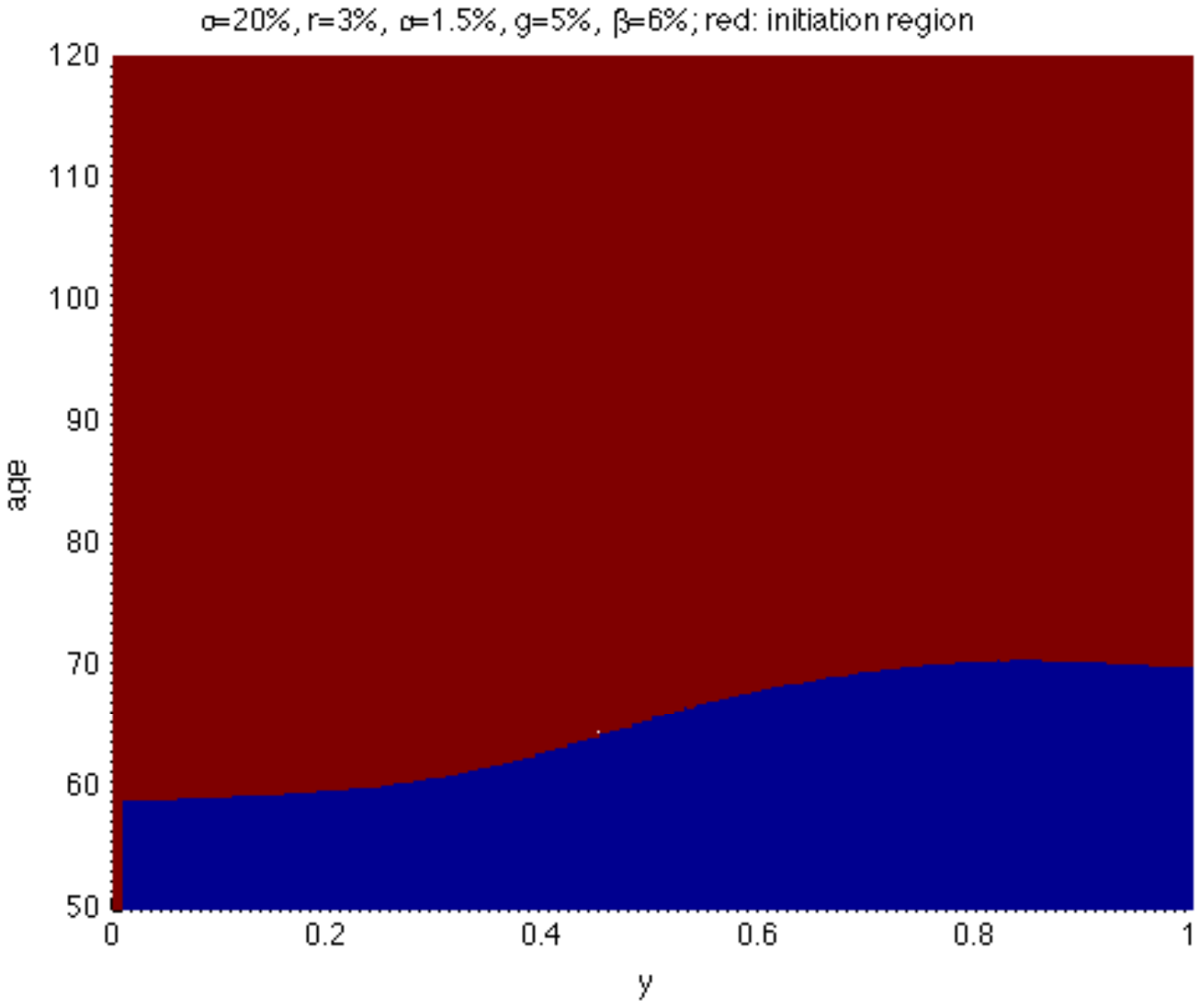} 
\vspace{-2in}
\caption{(High $\beta$) Model assumpions: $r=3\%$, $\beta=6\%$, $m=87.25$, $b=9.5$, $50\le\text{ age }\le120$, $\sigma=20\%$, $\alpha=1.5\%$, and $g=5\%$.}
\label{gcont1}
\end{center}
\end{figure}

\begin{figure}[here]
\vspace{-2in}
\begin{center}
\includegraphics[width=0.8\textwidth]{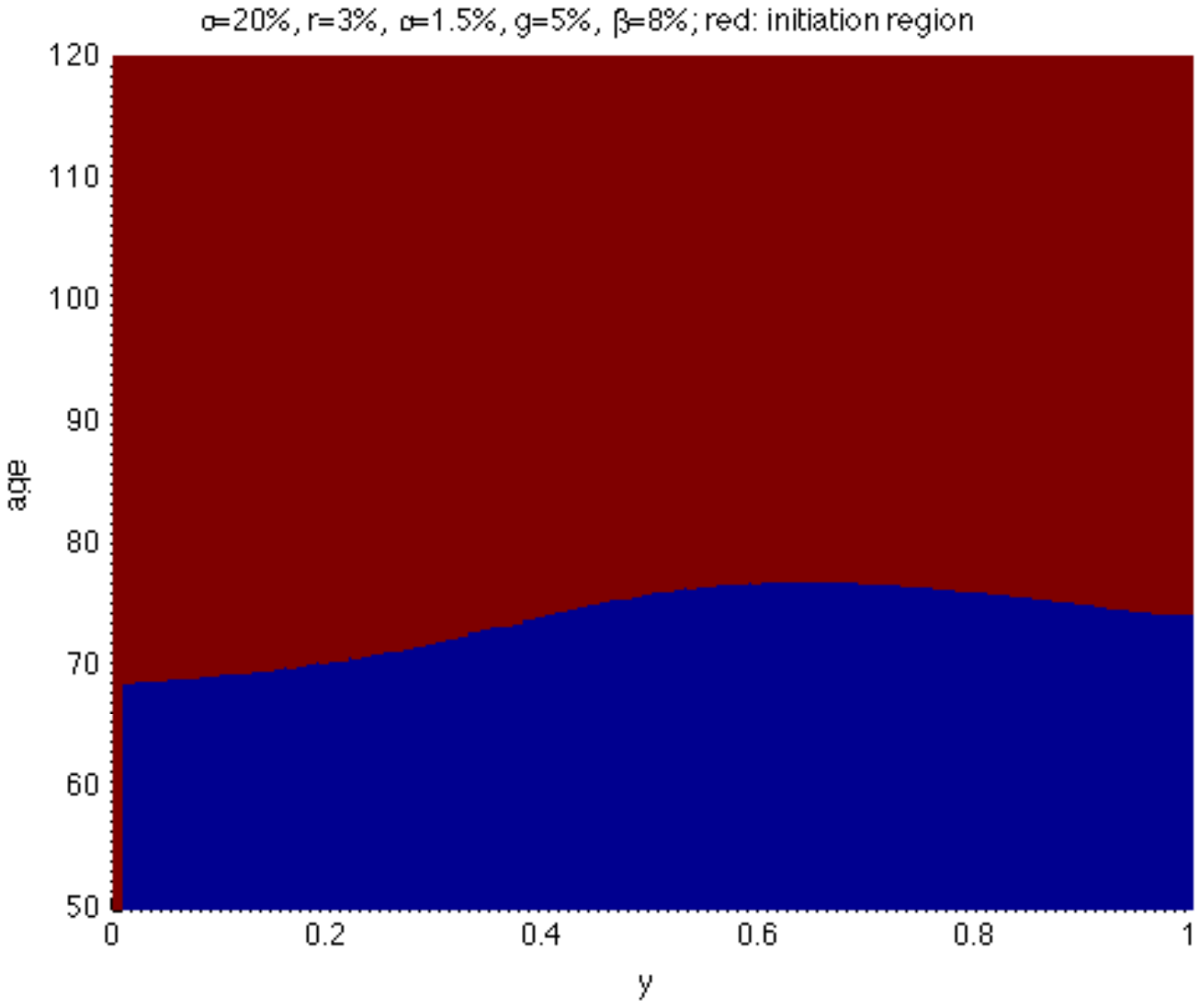} 
\vspace{-2in}
\caption{(Higher $\beta$) Model assumptions: $r=3\%$, $\beta=8\%$, $m=87.25$, $b=9.5$, $50\le\text{ age }\le120$, $\sigma=20\%$, $\alpha=1.5\%$, and $g=5\%$.}
\label{gcont2}
\end{center}
\end{figure}

\begin{figure}[here]
\vspace{-2in}
\begin{center}
\includegraphics[width=0.8\textwidth]{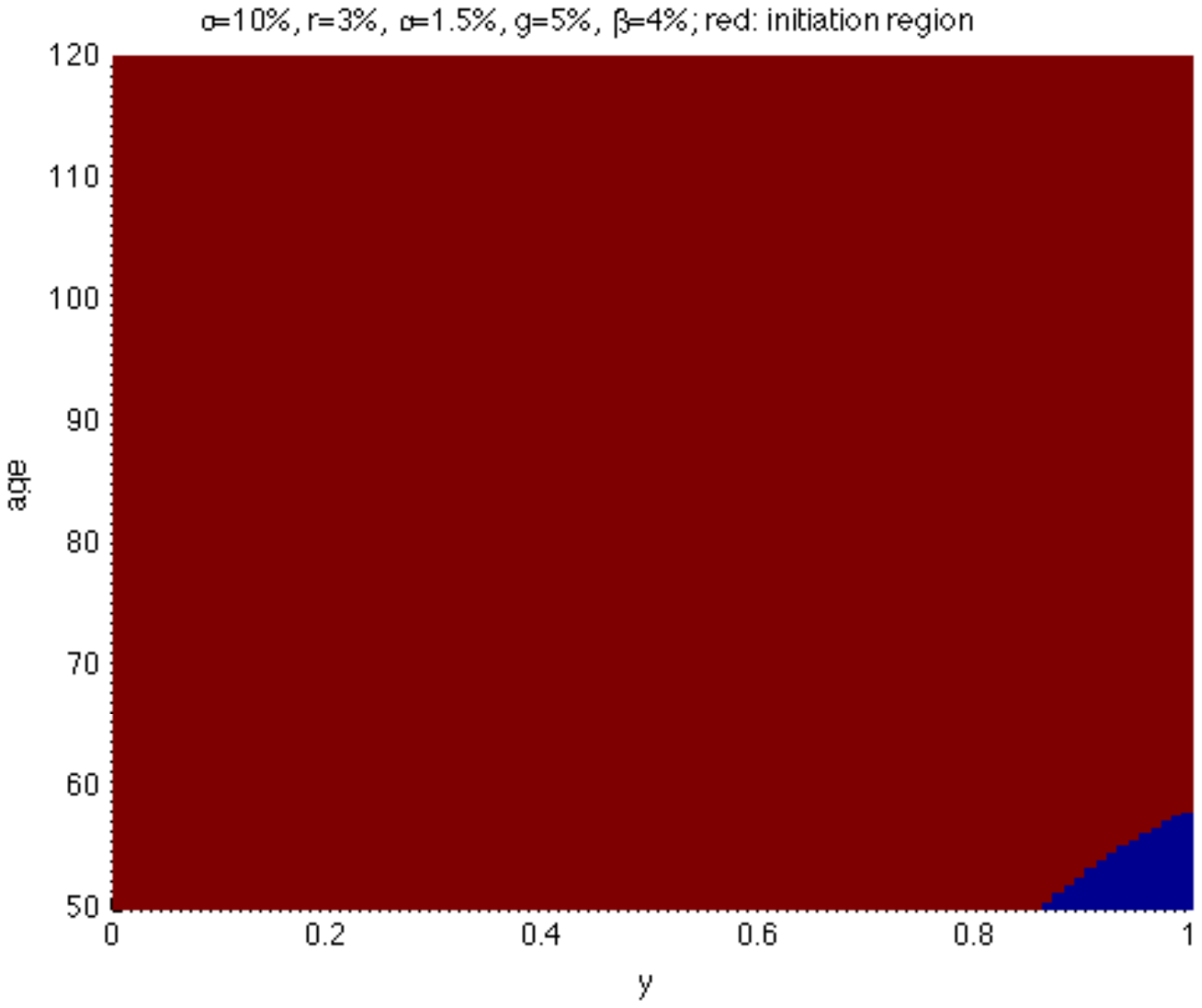} 
\vspace{-2in}
\caption{(Lower $\sigma$) Model assumptions: $r=3\%$, $\beta=4\%$, $m=87.25$, $b=9.5$, $50\le\text{ age }\le120$, $\sigma=10\%$, $\alpha=1.5\%$, and $g=5\%$.}
\label{gcont3}
\end{center}
\end{figure}

\begin{figure}[here]
\vspace{-2in}
\begin{center}
\includegraphics[width=0.8\textwidth]{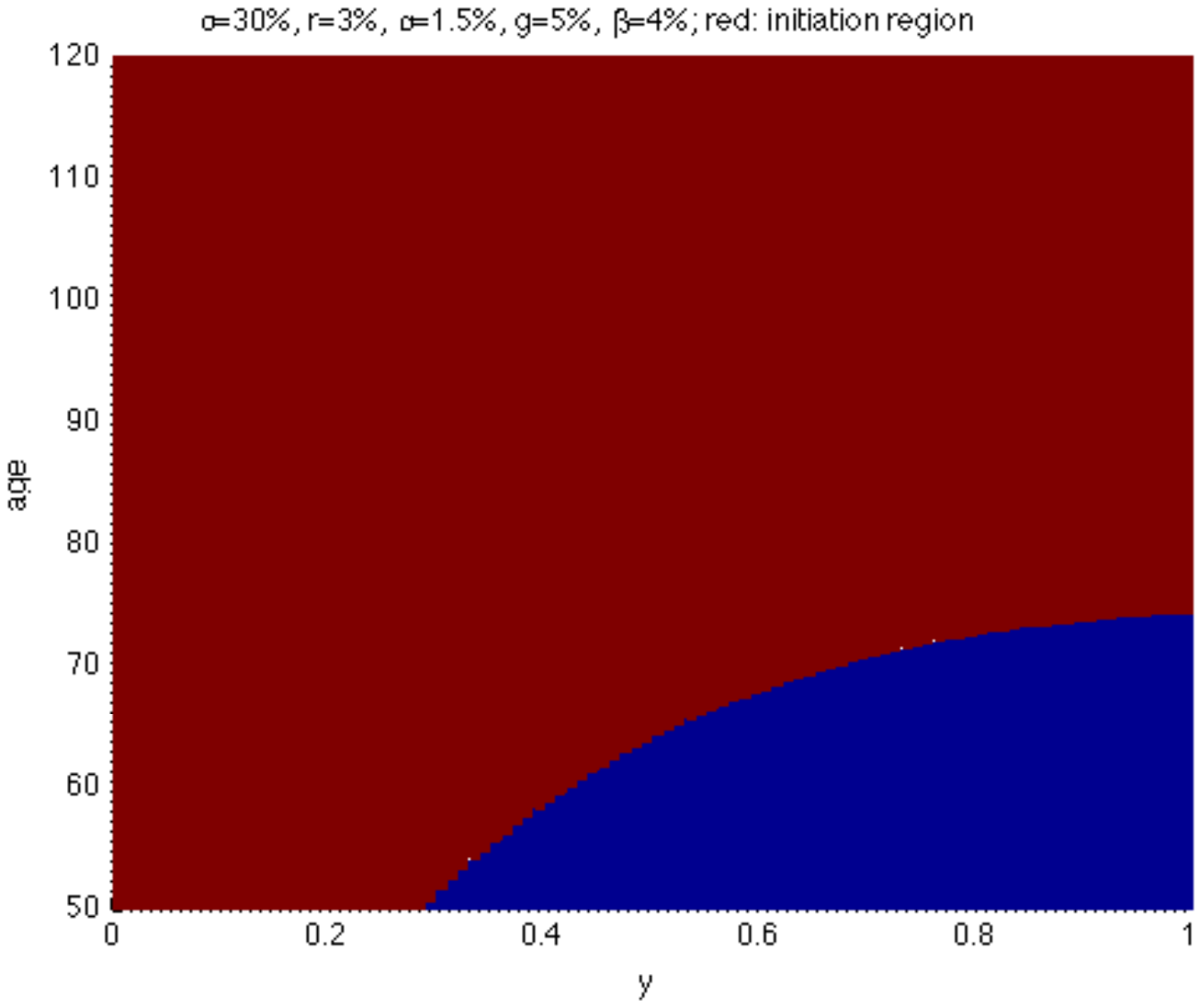} 
\vspace{-2in}
\caption{(Higher $\sigma$) Model assumptions: $r=3\%$, $\beta=4\%$, $m=87.25$, $b=9.5$, $50\le\text{ age }\le120$, $\sigma=30\%$, $\alpha=1.5\%$, and $g=5\%$.}
\label{gcont4}
\end{center}
\end{figure}

The effects of varying $\sigma$ (asset allocation restrictions) and $\beta$ (bonus and roll-up rates) are shown in Figures~\ref{gcont1}-\ref{gcont4}. Roughly speaking we find that the delay region expands as $\sigma$ or $\beta$ increases. Given that these products are intended as retirement savings vehicles, and that one will typically no longer contribute to a GLWB once withdrawals are initiated, the initiation question is typically asked by individuals in their late 50s and up. Even at a relatively high (for this type of investment) volatility of $\sigma=20\%$, as in Figure~\ref{gcont}, such individuals should delay only if the money-ness of the account approaches 1. At a more typical (and lower) volatility level of $\sigma=10\%$, as in Figure~\ref{gcont3}, there is no money-ness level at which delay is optimal for someone age 60. Situations at which delay is optimal for individuals at realistic ages require us to impose quite extreme parameter values, such as $\sigma=30\%$ (Figure~\ref{gcont4}) or a bonus rate of $\beta=6\%$ (Figure~\ref{gcont1}) or $\beta=8\%$ (Figure~\ref{gcont2}). All Figures~\ref{gcont1}-\ref{gcont4} assume the withdrawal rate $g$ is constant, regardless of whether the income is initiated at age 55 or age 85, for example.

Regardless of the exact parameter values, there are a number of qualitative insights that are immediately obvious after glancing at these figures. First, at older ages and lower values of money-ness ($y$) it optimal to stop accumulation and initiate the GLWB. This is the red region. The much smaller green region, which indicates that it is optimal to wait (and not initiate) only occurs in areas where the money-ness is very high and the age is quite young.

\begin{center}
\bf{Table \#1 Here}
\end{center}

Viewing the problem from a different angle, we can invert the model and compute the $\beta$ values needed for an individual to be indifferent between initiating immediately, versus initiating in five years for example, for several different ages and $y$ values. These values are reported in Table \#1. Notice how the threshold values are relatively higher than the (currently offered) 5\% or 4\% bonus for waiting. This is yet another indication of the optimality of initiation.

\begin{center}
\bf{Figure \#6 Here}
\end{center}

\begin{figure}
\vspace{-1.5in}
\begin{center}
\begin{tabular}{c}
\includegraphics[width=5in]{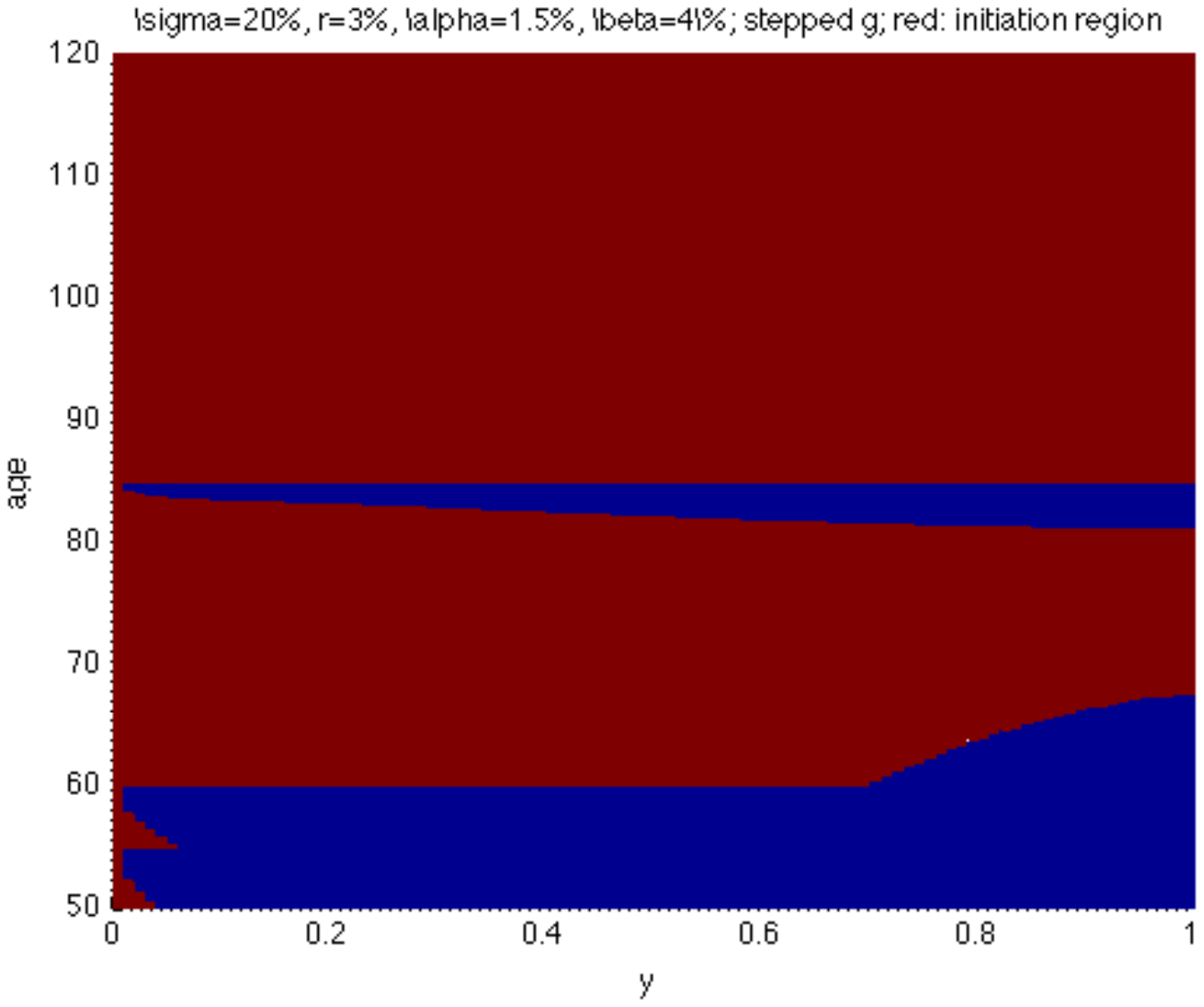} \\ 
\vspace{-4in}
\\
\vspace{-2in}
\includegraphics[width=5in]{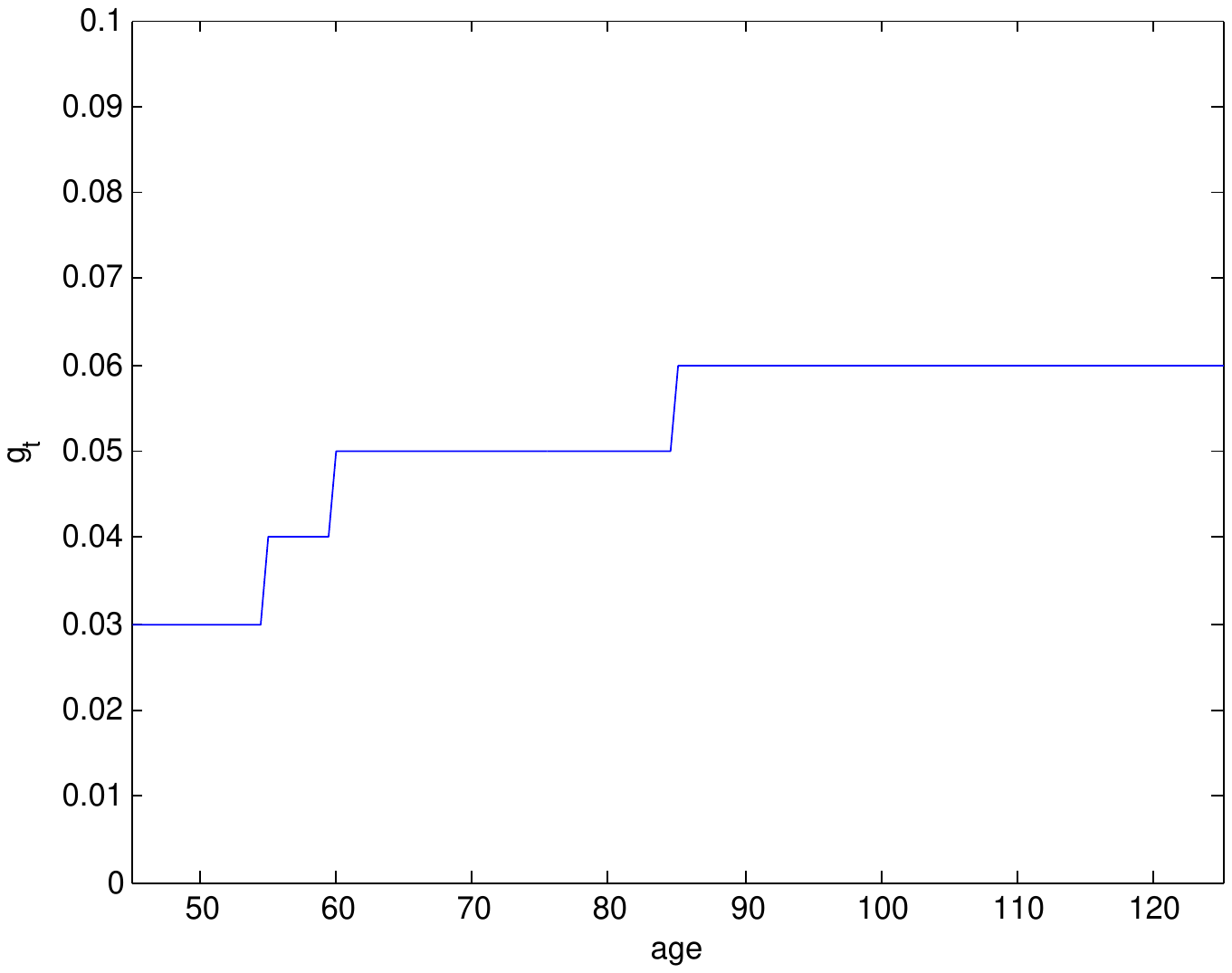} \\
\end{tabular}
\caption{(Stepped Payouts) Top part diplays the initiation (red) and non-initiation (green) regions. The bottom part displays the guaranteed payout rate $g_t$ as a function of initiation age. The model assumptions are that $r=3\%$, $\beta=4\%$, $m=87.25$, $b=9.5$, $50\le\text{ age }\le120$, $\sigma=20\%$, $\alpha=1.5\%$.}
\label{beta4gt}
\end{center}
\end{figure}

With a more realistic step function for $g_t$ which is dependent on age as in Figure \ref{beta4gt} (bottom) and using the same other parameter values as in Figure~\ref{gcont},  we see optimal initiation well below age 50, when the account is close to ruin. But, the boundary between the initiation and non-initiation regions becomes more complicated. 

Although Figure \ref{beta4gt} is based on a very specific payout rate function (g), the fragmented and non-contiguous region depicted is typical of most GLWBs and gets to the economic essence of the decision to initiate. Namely, there is a small sliver of age during which it is worth waiting to reach the next age band, and thus gain the higher lifetime income by immediately turning the GLWB on. The region in which the optimal policy changes from accumulate (green) to stop and turn-on (red) is driven entirely by the ages at which the guaranteed payout rate $g_t$ jumps to the next band. The thickness of the region depends on the money-ness of the account as well as the exogenous parameters such as the bonus (roll-up) rate, the interest rate and the volatility of the subaccounts. 

Finally, Table \#2 provides a summary of the comparative statics on the optimality of delay. The following parameter changes are all associated with greater propensity to initiate: (i.) higher ages and poor health, (ii) a bonus (roll-up) rate that is smaller, (iii.) a lifetime payout rate that is lower, (iv.) a volatility that is lower, or equivalently, an asset allocation within the variable annuity that contains less equity exposure, (v.) a risk-free valuation rate that is higher, (vi.) an insurance fee that is higher. Finally (vii.) the response to money-ness is more complicated. While it typically appears that lower values of money-ness are associated with initiation, there are some cases in which this is not the case. One should therefore be careful about concluding that all variable annuities that are (deeply) underwater should be initiated. It really depends on the bonus rate that is being offered to wait, as well as the lifetime payout rate itself.

\begin{center}
\bf{Table \#2 Here}
\end{center}

\subsection{Modeling Caveats}

There are a number of (subtle) modeling assumptions that we have made or have assumed in our analysis that are important to emphasize again, before these results are applied in practice.

First, we have essentially ignored the guaranteed minimum death benefit (GMDB) that is attached to most variable annuities (VA). Technically we assumed that at death the beneficiary receives the account value, only. Historically, most VAs offered a range of (lucrative) GMDB options in which the maximum account value could be protected at death, possibly with a guaranteed bonus (or roll-up) as well. These options (obviously) cost more and, more importantly, initiating withdrawals might reduce the GMDB in a disproportionate and undesirable manner. That said, one might question why a rational investor would select (and pay for) both the enhanced GLWB and the GMDB, since they protect against two opposing risks. Either way, our results are immediately applicable to individuals who have only elected the GLWB, which is a non-trivial portion of the variable annuity market.

Second, we have assumed that once withdrawals are initiated there are no further step-ups into higher income bands, in the lifetime payout amounts. For example, if the (deep in the money) GLWB was initiated at age 60, at a guaranteed base value of \$100,000 and a guaranteed payout rate (g) of 5\%, then the 5\% is paid for life even if $g$ at age 70 is $6\%$ In practice, some GLWB’s offer a further reset of the income amount into higher guaranteed payout rates as the account value reaches a new high-water mark after initiation. And, while the (risk-neutral) probability of this event occurring is quite small this is something we have not accounted for in the model. Nevertheless, we are confident that further resets even after initiation would only increase the initiation (red) region, all else being equal.

Third, while we have been careful to model the insurance fees as a percent of the guaranteed base ($M_t$) -- and not the account ($X_t$) value, which is a common oversight in the literature -- there is a baseline level of money management and insurance fees that is imposed on the account value ($X_t$), which we have (also) abstracted from. That said, this also creates an incentive to initiate earlier and help rush the account towards ruin.

Fourth, it is estimated that approximately half of GLWBs are sold with a joint-life rider in which the guaranteed payout rate ($g$) is reduced by 50 to 100 basis points in exchange for a life annuity that continues until the second death. Alternatively, some GLWBs charge higher fees for the same joint guaranteed payout rate. It is unclear how this affects the initiation decision, especially when there is a large age gap within the couple. Although, once again our comparative static analysis seems to indicate that reduced values of $g$ and/or increased fees $\alpha$ have the same effect. Namely, initiate.

In sum, given the increasing restrictions of sub-account volatility, where 100\% stock allocations are unheard of (so that sigma $<$ 15\%) and bonus rates and guaranteed payout rates are under 5\%, it is very difficult to justify paying for a GLWB rider and at the same time continuing the accumulation stage beyond the age of 60.

\section{Summary and Conclusion}

Variable Annuities (VA), also known as Segregated Mutual Funds in Canada -- contain a variety of living and death benefit (a.k.a. secondary) guarantees, which have attracted much attention within the insurance, economics and actuarial scholarly literature. Recently published articles have applied a No Arbitrage approach to the valuation of Guaranteed Minimum Death Benefits (GMDBs) and Guaranteed Lifetime Withdrawal Benefit (GLWB) policies; thus offering both a hedging and risk management perspective. This continues to be a fertile area of research and the results have attracted much attention from scholars, practitioners and the media. \footnote{In addition the papers listed in the literature review, see the recent work by Chen and Forsyth (2008), Donnelly, Jaimungal and Rubisov (2012), Huang and Forsyth (2011), Huang, Forsyth and G. Labahn (2012), Kling, Ruez and Russ (2011), Kolkiewicz and Liu (2012), Lin, Tan and Yang (2009), Marshall, Hardy and Saunders (2010), Marshall, Hardy and Saunders (2012), Piscopo and Haberman (2011) as well as Shah and Bertsimas (2008).} In this paper we continue this line of research by examining a No Arbitrage approach to optimal behaviour. 

One of the by-now established and stylized facts in this literature is that most (if not all) of the widely available GLWB products sold in the U.S. during the early part of the 21st century were severely underpriced.  Insurance companies and manufacturers were not hedging properly and/or they were using actuarial (as opposed to financial) techniques for pricing. Indeed, the popularity -- as well as the liability created by these products -- is responsible for the financial difficulties experienced by numerous insurance companies during the 2008/2009 crisis.  A number of large insurance carriers have recently exited this business, and the trend is expected to continue. In the current market environment, it is virtually impossible to locate VA + GLWB that provides the same level of value that was available just five years ago. In that sense, issuers and carriers have learned their lessons the hard way.

However, what has attracted less scholarly research attention but is equally important is the following question. How exactly should consumers who currently have (grandfathered) policies with these guarantees, optimize the value of their embedded option? In the U.S. market alone there are over \$1 trillion worth of these VA policies held by individuals. This tactical question is currently on the mind of many financial advisors with a large number of clients holding these products, in which the aggregate value of the embedded optionality is quite substantial.

From a financial economics perspective, this question is complicated by the fact that the optimal policy for a consumer seeking to maximize (and smooth) lifetime utility of consumption is not necessarily a policy that induces the maximum liability to the issuer. In other words, the hedging strategy (for the manufacturer) may not be the symmetric opposite of the dynamic utilization strategy (for the buyer).

Our results also indicate that it rarely makes sense to contribute additional funds to an existing VA + GLWB policy that is in-the-money. By contributing funds to an under-water policy the policyholder is watering-down the value of the guarantee. Quite perversely, consumers are unnecessarily doing the hedging for the insurance company. Much more importantly and quite relevant, we find that it is optimal to initiate the policy and begin withdrawals as soon as possible, with a few exceptions. When the payout rate is about to increase (very soon), the optimal policy may be to wait until the new payout rate kicks-in, and then initiate immediately.  Likewise, if the investor is unusually young, and able to benefit from a long lifetime of withdrawals at an elevated rate, it may be optimal to delay in order to accrue bonuses or stepups. Investors lucky enough to hold a GLWB with unusually generous terms (a very high bonus rate, or the ability to allocate significant funds to high-volatility assets) may find it optimal to delay. But for most investors, there is little reason to delay.

Even if an investor is still employed and/or is not interested in consuming the income from the GLWB, they are still better-off withdrawing the funds and re-investing in an equivalent (lower cost) mutual fund or tax-sheltered variable annuity. (This can be done via a partial 1035 exchange, in the language of the U.S. tax code.) And, if this ends-up ruining the account, so be it. They will have traded a GLWB for a non-guaranteed account plus a fixed life annuity.

Here is some intuition. The GLWB is worth something, only because of the probability the investment account will be (i) depleted by withdrawals at some date, and (ii.) the individual annuitant will live well beyond that date. The insurance is paid-for by ongoing charges to the account, which come to an abrupt end if-and-when the account hits zero. Thus, the sooner that account can be ``ruined'' and these insurance fees can be stopped, the worse it is for the insurance company and the better it is for the annuitant. Even the allure of a higher guaranteed base -- if you wait longer -- can't really offset the power that comes from depleting the account as soon as possible. As soon as the account is ruined, the annuitant and beneficiary are living off the insurance company's dime. And, while it might seem odd that trying to increase the hedging cost to the insurance company is in the best interest of the policyholder, our utility-based analysis indicates a similar phenomenon. 

We caution that our research is ongoing, and these results are dependent on modeling assumptions in which market volatility and long-term interest rates remain at their current level. If, for example, the VIX index jumps to elevated levels for an extended period of time and remains elevated, and/or long-term bond yields return to their historical levels, some of these high-level results may no longer be valid.

\newpage

\newpage

\newpage

\newpage

\newpage

\newpage

\newpage

\begin{table}[htb*]
\begin{center}
\caption{Notes: The PDE assumes a continuous (step-up, interest, return) model versus MC which assumes yearly adjustments. The initiation is after a five year delay regardless of the moneyness value.  Note that the higher the current age, the greater the break-even bonus must be to justify {\bf not} initiating. Also, if the volatility of the sub-accounts is lower than the $\sigma=20\%$ assumed, or the insurance fee is greater than the $\alpha=1.5\%$ asssumed or the guaranteed payout rate is less than the $g=5\%$ assumed, the resulting break-even $\beta$ values would be higher than reported. These results should also provide an indication of the error introduced by using continuous versus annual compounding.\label{table1}}
\begin{tabular}{||c||c||c||c||}
\hline\hline
\multicolumn{4}{||c||}{\bf{Bonus Rate That Will Induce Waiting}}\\
\hline
\multicolumn{4}{||c||}{\emph{PDE: bonuses/fees/stepups accrue continuously}}\\
\hline
\multicolumn{4}{||c||}{\emph{MC: bonuses/fees/stepups accrue yearly}}
\\ \hline\hline
Current Age & Moneyness & PDE bonus & MC bonus  \\ \hline\hline
Age 55 & $y=1.0$ & $\beta=3.62\%$ & $\beta =3.97\%$ \\ 
\hline
Age 55 & $y=0.8$ & $\beta=4.15\%$ & $\beta =4.53\%$ \\ 
\hline
Age 55 & $y=0.5$ & $\beta=5.30\%$ & $\beta =5.29\%$ \\ 
\hline\hline
Age 65 & $y=1.0$ & $\beta =5.78\%$ & $\beta =5.99\%$ \\ 
\hline
Age 65 & $y=0.8$ & $\beta =5.98\%$ & $\beta =6.13\%$ \\ 
\hline
Age 65 & $y=0.5$ & $\beta =6.74\%$ & $\beta =6.69\%$ \\ 
\hline\hline
Age 75 & $y=1.0$ & $\beta =9.19\%$ & $\beta =9.36\%$ \\ 
\hline
Age 75 & $y=0.8$ & $\beta =8.92\%$ & $\beta =8.84\%$ \\ 
\hline
Age 75 & $y=0.5$ & $\beta =9.25\%$ & $\beta =9.12\%$ \\ 
\hline\hline
\end{tabular}
\smallskip
\end{center}
\end{table}

\newpage

\begin{table}[htb*]
\begin{center}
\caption{Notes: The table displays the directional change in underlying parameter variables that would result in initiation of the GLWB assuming one is currently on the razor's edge between the two regions. \label{table2}}
\begin{tabular}{||c||c||}
\hline\hline
\multicolumn{2}{||c||}{\bf{Comparative Statics for Variables that Affect GLWB
Initiation}} \\ \hline\hline
Age or Gender or Health $(\lambda _{t})$ $\uparrow $ & Older =
Initiate \\ \hline\hline
Waiting Bonus = Roll-Up Rate $(\beta )$ $\downarrow $ & Smaller = Initiate \\ \hline\hline
Lifetime Income Payout Rate $(g_{t})$ $\downarrow $ & Reduced = Initiate \\ \hline\hline
Stock Allocation $(\sigma )$ $\downarrow $ & Restricted =
Initiate \\ \hline\hline
Risk-free Treasury Rate $(r)$ $\uparrow $ & Increase = Initiate
\\ \hline\hline
Insurance Fee as \% of Base $(\alpha )$ $\uparrow $  & More Expensive = Initiate \\ \hline\hline
Account Value $\div $ Guarantee $(y)$ $\downarrow $ & Underwater* = Indeterminate \\ \hline\hline
\end{tabular}
\smallskip
\end{center}
\end{table}

\end{document}